\documentclass[prb,preprint]{revtex4-1}

\usepackage{amsmath}  % needed for \tfrac, \bmatrix, etc.
\usepackage{amsfonts} % needed for bold Greek, Fraktur, and blackboard bold
\usepackage{graphicx} % needed for figures
\usepackage[normalem]{ulem}
\usepackage[usenames,dvipsnames]{color}

\newcommand\uv[1]{\textquotedblleft #1\textquotedblright}

\begin{document}

\title{The Role Of Dimensional Analysis In Teaching Physics}

\author{Michaela Reichelova}
\email{michaela.reichelova@ukf.sk} % optional
% optional second address
% If there were a second author at the same address, we would put another
% \author{} statement here.  Don't combine multiple authors in a single
% \author statement.
\author{Aba Teleki}
\email{ateleki@ukf.sk} % optional
\affiliation{Department of Physics, Constantine the Philosopher University, SK-94974, Nitra, Slovakia}
% Please provide a full mailing address here.

\date{\today}

\begin{abstract}
Dimensional analysis is a simple qualitative method for determining essential connections between physical quantities. It is applicable to a multitude of physics problems, many of which can be introduced early on in a university physics curriculum. Despite the relative simplicity of the approach, it is rarely included in physics curricula. Here, we apply dimensional analysis to find the radiated power from an electric dipole. Employing dimensional analysis, we demonstrate a straightforward way to derive the latter without the need for complex mathematical treatments.

\end{abstract}

\maketitle
\section{Introduction}
Dimensional analysis is a method of working with physical quantities and their physical dimensions. The approach can be used not only to estimate typical values of various quantities, but also to verify predicted relations between elements of a theory.
Edgar Buckingham wrote an exhaustive article on this topic 100 years ago.\cite{Buckingham1914} Many comprehensive textbooks also exist on dimensional analysis (see \cite{Bridgman1922} for example).

Dimensional analysis is rarely a subject of secondary school or early university physics curricula. This is rather unfortunate, because dimensional analysis plays an important role in the development of qualitative thinking and in the recognition of similarities in systems.%
%
% Begin footnote
\footnote{ Bao~\cite{Bao2009} suggests that rigorous training of physics knowledge in high schools has made a significant impact on students' ability to solve physics problems, while such training does not seem to have direct effects on their general ability at scientific reasoning. Marusic~\cite{Marusic2011} shows that traditionally formulated numerical exercises, which are frequently used in teaching physics, do not develop students' abilities in physical thinking to a desirable extent, and thus fail to fulfill one of the most important aims of teaching physics. Blasiak~\cite{Blasiak2012} states that comprehension in physics is a complex and multi-stage process. It is relatively easy to describe a phenomenon, but more difficult to explain it, whereas predicting it, which in fact is the crowning achievement of adequate comprehension, seems to be the most difficult task of all.}
% End footnote

Many studies indicate that physical phenomena are more deeply understood if the interpretation is firstly qualitative, and describes the physical properties that are most important in a phenomenon. It is useful if a quantitative description in terms of formulas follows such an introduction. Taber~\cite{Taber2009} and Ploetzner~\cite{Ploetzner1999} have found that \uv{quantitative information is likely to be more easily integrated into qualitative problem representations than the other way around.}
In our opinion, dimensional analysis is a suitable intermediate step in solving problems that are more complex. However, the question of complexity is age-dependent. Dimensional analysis is particularly suitable in cases where the mathematical complexity of the problems to be solved is too high for the age of the given pupils or students. Also, in his Millikan Lecture 2009 Eisenkraft~\cite{Eisenkraft2010} emphasizes that we "should develop teaching strategies that enable us to share an understanding of physics with all students because everyone deserves an opportunity to reflect on the wondrous workings of our universe". Dimensional analysis could make physics accessible to students with lesser mathematical skills. Dimensional analysis is also suitable in cases where the mathematical complexity outweighs the benefit of the calculations. For example, this facet of dimensional analysis had been used to derive the gravitational power radiated by a celestial body that moves in a circular orbit,\cite{Bracco2009} the resistance force of a fluid that occurs when a body moves through it,\cite{Misic2010} and the speed of the propagation of waves on water.\cite{Misic2010} The number of works in the recent literature~\cite{Pescetti2008,Pescetti2009,Aghamohammadi2011,Bernal2013,Perkalskis2004} attest to the enormous utility of this method in various disciplines of physics.

An interesting phenomena
is the radiation emanated by an electric dipole. \footnote{The analogy between electric and magnetic dipoles is straightforward,\cite{Bezerra2012} and one can repeat our steps to derive the power radiated by a magnetic dipole as well. The formula for the power radiated by the dipole is derived in several publications in various ways.\cite{Likar2009,Jackson1998,Jefimenko1989,Sedlak2002} Jackson~\cite{Jackson1998} is a classic reference.}

While evaluation of the power radiated by a dipole is one of the exactly solvable classical electromagnetic problems, solving it still requires multivariable calculus. An alternative approach is therefore desirable for both the non-calculus based courses and for the exposition of the subject at the high-school level.
At first glance it may appear that taking the planetary model and applying dimensional analysis to calculate the power radiated by an electron encircling the nucleus of an atom on a circular trajectory could work. Unfortunately, the system of equations that this leads to is under-determined, and the dimensional analysis fails (see discussion after Eq.~(\ref{e:09})).
In what follows, we avoid this difficulty by applying dimensional analysis first to a more general problem without fixing any concrete geometry. Subsequent application to specific cases then only requires finding a factor of the order of one.  This derivation is very simple, and it does not require a high level of mathematical skill. The only requirement with respect to the student's prior knowledge is the ability to solve a simple system of linear equations.

Finally, the calculation of the radiated power for the planetary model of an atom in non-calculus-based courses is not an art for art's sake. This problem is closely connected with the argumentation for why the planetary model of the atom was rejected even before Rutherford's famous scattering experiment~\cite{Campbell999} and Bohr's model of the atom.\cite{Heilbron1981} It is exciting for students to understand, quantitatively and otherwise, that classical electromagnetism predicts strongly radiating electrons in atoms, which is in contradiction to everyday experience.

\section{The radiated power of an accelerating electric charge}

How intense is this radiation? Is it really so important that the revision of the fundamental principles of physics is required? Can this question be resolved by using dimensional analysis?

When a student is faced with this problem (in the topic of dimensional analysis), it is natural for him to express the radiated power using quantities such as the radius $r_0$ of the atom, the mass $m$ of the electron, etc., which describe the properties of the planetary model. And this is the big difference if one includes $m$ in the list of variables. In this case the solution is not unique. Our experience is that students  faced with the problem of planetary model do not think over the general case. They include the mass in the list of variables automatically. It may be a surprise for beginners that a more general problem (as in our case) can lead to a unique solution while the special case does not. Selecting the appropriate variables is, after all, the hardest part of dimensional analysis. Solving the linear equations connected with the question is the easy part.

A typical source of dipole radiation is a particle with electric charge undergoing simple harmonic motion. To derive its total radiated power $P$ in a vacuum, we will assume a general case, namely a particle with electric charge $q$ and with an arbitrary acceleration $a.$

Charged particles do not generate electromagnetic radiation while at rest ($v=0$ and $a=0$), and therefore a particle with a constant velocity $\vec{v},$ does not, either. Relativity forbids this, because inertial frames are equivalent. In this case, the co-moving frame is an inertial frame -- it is moving with the charged particle. In the co-moving inertial frame, the particle is at rest and does not generate electromagnetic radiation. Accelerated particles are a different case. Accelerated charged particles (electrons) can be found, for example, in a dipole antenna (e.g. in a mobile phone) -- the alternating voltage in the electric circuit accelerates electrons in the circuit.

Such an accelerated particle generates electromagnetic radiation. Since this radiation is an electromagnetic phenomenon, the radiated power $P$ will also depend on the electrical and magnetic characteristics of the vacuum: vacuum permittivity $\varepsilon_0$ and vacuum permeability$\mu_0.$ Because these two quantities determine the speed $c$ of electromagnetic waves in a vacuum according to the relation $c=1/\sqrt{\varepsilon_0\mu_0},$ we have decided to replace the vacuum permeability $\mu_0$ with the speed of light in a vacuum $c,$ because students are more familiar with the latter quantity. Acceleration $\vec{a}$ is a vector quantity, but we are interested in the total power $P$ radiated by the electric charge of the particle in all directions, which is a scalar. This leads us to assume that (due to the integration in all directions) there will only be dependence on the magnitude of acceleration $a.$ There is no need to actually mention this reasoning to students, but it should be thought of in case questions arise about the vector nature of acceleration.  We further assume that the radiated power does not depend on the mass of the moving charge, because the source of the electromagnetic field is the charge of the particle and not its mass.%
%
% begin footnote
\footnote{A particle with no electric charge and no magnetic dipole has no electric field of its own, and its acceleration does not generate electromagnetic waves (electromagnetic radiation).} %
% end footnote
%
Therefore the radiation is the same for protons and electrons undergoing the same acceleration.

In general, the radiated power $P$ is a function of the aforementioned quantities, i.e.,
\begin{equation}\label{corr:01}
  P=f(q,c,\varepsilon_0,a).
\end{equation}

It is necessary for the physical dimensions $[P],$ $[q],$ $[c],\dots$ of the quantities $P,q,c,\dots $ to satisfy this equation.

The first question is whether one can express the dimensions of $[P]$ as a combination of dimensions of the quantities on the right-hand side of Eq.~(\ref{corr:01}), i.e., do real numbers $\alpha,\beta,\gamma,\delta$ exist such that
\begin{equation}\label{corr:02}
  [P]=[q]^\alpha [c]^\beta [\varepsilon_0]^\gamma [a]^\delta?
\end{equation}

We emphasize here that physical consideration plays a fundamental role in finding the appropriate quantities that describe a concrete physical system. It is very important for beginners (e.g. in secondary schools or undergraduate courses), to apply dimensional analysis to problems with a unique solution (as in our case). In many cases we can also simplify physical assumptions, realizing that the physically important variable does not depend on certain variables in the discussed range of importance. Physical insight is needed to determine the appropriate significant variables for the scale of the problem, e.g., the surface tension of water is irrelevant to the propagation of large amplitude waves. This is why dimensional analysis is so beneficial for undergraduate students too.

In this spirit, mathematically, the radiated power $P$ is given by the following expression:

\begin{equation}\label{e:01}
P = Kq^\alpha c^\beta {\varepsilon_0}^\gamma a^\delta,
\end{equation}
where $K$ is a dimensionless constant. For the dimensions of power $P,$ we write $[P]=ML^2 T^{-3};$ for the dimensions of the speed of light $c,$ we write $[c]=LT^{-1};$ for the dimensions of acceleration $a,$ we write $[a]=LT^{-2};$ for the dimensions of charge $q,$ we write $[q]=IT;$ and for the dimensions of permittivity of the vacuum $\varepsilon_0,$ we write $[\varepsilon_0]=I^2 T^4 L^{-3} M^{-1}.$ Here, $I,T,L$, and $M$ are the dimensions of the basic physical quantities of current, time, length, and mass, respectively. From the equality of the right-hand and left-hand sides we get
\begin{equation*}
ML^2 T^{-3} = (IT)^\alpha (LT^{-1})^\beta (I^2 T^4 L^{-3} M^{-1})^\gamma (LT^{-2})^\delta.
\end{equation*}

Simplifying the equation we obtain
\begin{equation}\label{e:02}
ML^2 T^{-3} = I^{\alpha + 2\gamma} T^{\alpha - \beta + 4\gamma - 2\delta} L^{\beta - 3\gamma + \delta} M^{-\gamma}.
\end{equation}
From the equality of the right-hand and left-hand sides we obtain four equations, each expressing the equality for one of the exponents $\alpha,$ $\beta,$ $\gamma,$ $\delta$ in individual dimensions
\begin{eqnarray}
M&:&\phantom{-} 1=-\gamma,\nonumber\\
L&:&\phantom{-} 2=\beta+\delta-3\gamma,\nonumber\\
T&:&-3=\alpha-\beta+4\gamma-2\delta,\nonumber\\
I&:&\phantom{-} 0=\alpha+2\gamma.\nonumber
\end{eqnarray}
After solving the equations we get $\alpha=2$, $\beta=-3$, $\gamma=-1$ and $\delta=2.$ Thus, for the radiated power $P$ we must write
\begin{equation}\label{e:03}
P=K\frac{q^2 a^2}{\varepsilon_0 c^3}.
\end{equation}

\section{Discussion}
Dimensional analysis is a very suitable tool for teaching pupils in the lower grades where dimensional tests are already known and required as a part of certain problem. Also, students in the lower grades should be able to solve a system of linear equations. There are many simple but interesting physical phenomena solvable by dimensional analysis: the period of a simple pendulum, the speed of sound in a string or in fluids and gases, the phase speed of waves on a surface of fluids, and many others.

On one hand, dimensional analysis cannot determine the value of dimensionless constant $K,$ on the other hand, gives a rigorous answer to questions like (for example) \emph{what happens with the radiation power $P,$ if one increase the acceleration $a$ by factor 2} (doubling $a$ quadruples $P$).

In the case of basic physical phenomena, dimensionless constant $K$ can be determined from experiment, as in, for example, the period of a mathematical pendulum. In a general case one can assume that $K$ represents certain geometric properties of the problem, and therefore its magnitude is of the order of one -- this is an acceptable approximation if there is no possibility to obtain a more precise estimation.

To confirm the correctness of our arguments, it may be interesting to obtain the constant $K$ by comparing results in the literature. Our derivation says nothing about the nature of acceleration $a$, which affects the value of the as-of-yet undetermined constant $K$. The two simplest-to-analyze cases are:  (1) a particle undergoing simple harmonic motion, and (2) a particle in a uniform circular motion. Another possible assumption could be that of a constant acceleration ($a=const.$), which holds for Bremsstrahlung and collisions of two particles (in the simplest approximation).

In the case of simple harmonic motion in one direction, we can determine the constant $K$ from Eq.~(\ref{e:04}) which was derived by Jefimenko~\cite{Jefimenko1989} (see \S 16-8, page 561)
\begin{equation}\label{e:04}
P=\frac{{p_0}^2 \omega^4}{12\pi\varepsilon_0 c^3}=\frac{1}{4\pi \varepsilon_0}\frac{{p_0}^2 \omega^4}{3c^3},
\end{equation}
where $p_0$ is the dipole moment ($p_0=qr_0$ and $a_0=\omega^2r_0$ where $r_0$ is the amplitude of oscillation) and $\omega$ is the angular velocity.

In the case of a uniform circular motion of a particle in one plane (with acceleration $a=r\omega^2,$ where $r$ is the radius of the trajectory), we determine the constant $K$ from the relation derived by Jackson \cite{Jackson1998} (see \S 14.2, page 665). In the SI system it has the form of
\begin{equation}\label{e:08}
P=\frac{2}{3}\frac{1}{4\pi \varepsilon_0}\frac{q^2 a^2}{c^3}.
\end{equation}

These relations (\ref{e:04}) and (\ref{e:08}), can be brought to the same form of
\begin{equation}\label{e:06}
P=\frac{q^2 a^2}{6\pi\varepsilon_0 c^3}.
\end{equation}

For harmonic motion (see Eq.~(\ref{e:04})) $a$ denotes the root mean square (rms) acceleration ($a=a_0/\sqrt{2},$ where $a_0$ is the maximum acceleration).

From Eq.~(\ref{e:06}) we can determine constant $K,$ which has a value of  $K=\frac{1}{6\pi}\approx 0.053.$ This value of $K$ is not close to unity.

One can appreciate that in SI units (used here), $\varepsilon_0$ is mainly accompanied by the geometric factor $4\pi$ (this is the consequence of the form of Maxwell's equation $\nabla \cdot {\mathbf{E}} =\rho/\varepsilon_0$ -- in a vacuum).  Taking this fact into consideration, one obtains the formula for the radiated power, Eq.~(\ref{e:03}), in the form
\begin{equation}\label{e:09}
P=K'\frac{q^2 a^2}{4\pi \varepsilon_0 c^3}.
\end{equation}
Now, we determine $K'=\frac{2}{3},$ and it is in perfect accordance with our statement about the magnitude of the constant $K'$.

The radiation of the electron revolving in the classical atom model is an important part leading to the problem of quantization. The classical argument against the planetary atom can be found in most introductory textbooks: \uv{The hydrogen atom possesses only a single electron, which revolves round the nucleus. By the rules of electrodynamics, an  electron accelerated like this sends out radiation continuously, and so loses energy; in its orbit it would therefore necessarily get nearer and nearer the nucleus into which it would finally plunge.} (see \cite{Born1969} page 81).

Our experience is, that students assume the radiated power $P$ in the form  (planetary model) $P=f(e,c,\varepsilon_0,m,r_0)=K e^\alpha c^\beta \varepsilon_0^\gamma m^\delta r_0^\phi,$
where $e$ is the elementary charge, $m$ is the mass of the electron and $r_0$ is the radius of circular trajectory of the electron in the hydrogen atom (Bohr radius).
Parameters $\alpha,\beta,\dots$ do not have a unique solution. We have one parametric class of solutions $\alpha=-2(1+\phi),$ $\beta=5+2\phi,$ $\gamma=1+\phi$ and $\delta=2+\phi;$ we choose $\phi$  as a parameter. We need an additional physical argument to assign the parameter $\phi$; in other words, the problem based on our assumptions is under-determined. Selecting the appropriate variables is, after all, the hardest part of dimensional analysis. Solving the linear equations is the easy part.

\textbf{Note:} Permeability $\varepsilon$ and especially the permeability of the vacua $\varepsilon_0$ is not a dimensionless quantity in SI. This is a consequence of the definition of units in SI, namely units connected with electricity as electric charge $q,$ electric current $I,$ the intensity of electric field $\mathcal{E},$ and so on. Combining them with the permeability of vacua $\varepsilon_0$ in an appropriate form (in the form $q^2/(4\pi\varepsilon_0), I/\sqrt{4\pi\varepsilon_0}, \sqrt{4\pi\varepsilon_0}\,\mathcal{E},\dots$) we can eliminate the unit \emph{ampere} in every case. We decrease, in this way, the number of parameters and also the number of independent equations with one, simultaneously. We recommend to repeat our derivations, demonstrated here, with this simplification assuming the radiated power in the form $P=(q^2/(4\pi\varepsilon_0))^\alpha c^\beta a^\delta$.

\section{Conclusion}
The aim of this article is to point out the usefulness of a simple qualitative method: dimensional analysis. We have calculated the radiated power generally for an accelerated charged particle. A particular case of electric dipole radiation is also discussed. The derivation shown above has been used with success by the authors in introductory courses of quantum physics and nuclear physics, in particular to argue for the quantum properties of atoms.  The exact solution Eq.~(\ref{e:08}) takes for circular trajectory of the electron in hydrogen atom following form
\begin{equation}\label{e:07}
P=\frac{2}{3}\frac{1}{(4\pi\varepsilon_0)^3}\frac{e^6}{m^2 c^3 r_0^4},
\end{equation}
where $e=1.602\times 10^{-19}\,\mathrm{C}$, $\varepsilon_0=8.854\times 10^{-12}\,\mathrm{F}\cdot\mathrm{m}^{-1}$, $r_0=5.292\times 10^{-11}\,\mathrm{m}$, $m=9.109\times 10^{-31}\,\mathrm{kg}$, $c=2.998\times 10^8\,\mathrm{m}\cdot\mathrm{s}^{-1}$.\cite{CODATA2010} By using Eq.~(\ref{e:07}) one can calculate that a classical approach gives the radiated power of the electron in the hydrogen atom as $P\approx10^{-8}\ \mathrm{W}$. Should this be the actual radiated power, a human body with its $10^{28}$ electrons would emit power comparable to the peak output of a nuclear bomb. The radiated power can be converted into a lifetime for the hydrogen atom using the conservation of energy and it gives $t\approx10^{-11}\,\mathrm{s}.$ These findings are clearly in contradiction to everyday experience. Dimensional analysis is an especially useful method for students, also for those with strong mathematical skills.
One can say that dimensional analysis is a form of the Fermi problem-solving strategy, because it naturally leads the student to think about the physical nature of the problem. As in the Fermi approach, the dimensional method breaks the problem into a sequence of steps that are easier for students to follow, namely the definition of the physical model; making assumptions -- determining variables and parameters; using simple mathematics to compile equations and solving them, which finally leads to an approximate solution.

\begin{acknowledgments}
This research is a part of a wider effort aimed at using Fermi problem solving as a background for classroom exposition of essential physical phenomena. The work was supported by several grants: KEGA UKF61-4/2012, PRIMAS-Promoting Inquiry in Mathematics and Science Education Across Europe. Funding body: The European Union represented by the European Commission. Funding scheme: Coordination and support action. Call FP7-SCIENCEIN-SOCIETY-2009-1. Grant agreement number: 244380 and A-CENTRUM (ITMS 26110230026). The authors are grateful for their useful discussions with M. Kolesik and R. Durny.
\end{acknowledgments}
%
%
%\bibliography{References}
%\bibliographystyle{aipnumDipole}

%

\end{document}